\documentclass[sn-mathphys,Numbered]{sn-jnl}
\usepackage{a4wide}
\usepackage{mathpazo}
\usepackage{graphicx}%
\usepackage{multirow}%
\usepackage{amsmath,amssymb,amsfonts}%
\usepackage{amsthm}%
\usepackage{mathrsfs}%
\usepackage[title]{appendix}%
\usepackage{xcolor}%
\usepackage{textcomp}%
\usepackage{manyfoot}%
\usepackage{booktabs}%
\usepackage{algorithm}%
\usepackage{algorithmicx}%
\usepackage{algpseudocode}%
\usepackage{listings}%
\usepackage{adjustbox}
\usepackage{ulem}
\usepackage{paracol}

\begin{document}
\title{
Mapping Regional Disparities in Discounted Grocery Products 
}
\author[a,b,+,$*$]{\fnm{Antonio} \sur{Desiderio}}
\author[c,d,+]{\fnm{Alessia} \sur{Galdeman}}
\author[a,+]{\fnm{Franziska} \sur{Bäuerlein}}
\author[a,e,$*$]{\fnm{Sune} \sur{Lehmann}}
\affil[a]{\orgdiv{Department of Applied Mathematics and Computer Science}, \orgname{Technical University of Denmark}, \orgaddress{\street{Richard Petersens Plads}, \postcode{2800}, \state{Copenhagen}, \country{Denmark}}}
\affil[b]{\orgdiv{ISI Foundation}, \orgaddress{\street{Via Chisola 5}, \postcode{10126}, \state{Turin}, \country{Italy}}}
\affil[c]{\orgdiv{Data Science Section}, \orgname{IT University of Copenhagen}, \orgaddress{\street{Rued Langgaards Vej 7}, \postcode{2300}, \state{Copenhagen}, \country{Denmark}}}
\affil[d]{\orgdiv{Computer Science Department}, \orgname{University of Milan}, \orgaddress{\street{via Celoria 18}, \postcode{20133}, \state{Milan}, \country{Italy}}}
\affil[e]{\orgdiv{Copenhagen Center for Social Data Science (SODAS)}, \orgname{University of Copenhagen}, \orgaddress{\street{Øster Farimagsgade 5}, \postcode{1353}, \state{Copenhagen}, \country{Denmark}}}
\affil[$+$]{These authors contributed equally to this work}
\affil[$*$]{Corresponding authors: \href{mailto:antde@dtu.dk}{antde@dtu.dk} and \href{mailto:sljo@dtu.dk}{sljo@dtu.dk}.}
\abstract{
Food waste represents a major challenge to global climate resilience, accounting for almost 10\% of annual greenhouse gas emissions.
The retail sector is a critical player, mediating product flows between producers and consumers, where supply chain inefficiencies can shape which items are put on sale. 
Yet how these dynamics vary across geographic contexts remains largely unexplored.
Here, we analyze data from Denmark’s largest retail group on near-expiry products put on sale. 
We uncover the geospatial variations using a dual-clustering approach.
We characterize multi-scale spatial relationships in retail organization by correlating store clustering -- measured using shortest-path distances along the street network -- with product clustering based on promotion co-occurrence patterns.
Using a bipartite network approach, we identify three regional store clusters, and use percolation thresholds to corroborate the scale of their spatial separation.
We find that stores in rural communities put meat and dairy products on sale up to 2.2 times more frequently than metropolitan areas. 
In contrast, metropolitan and capital regions lean toward convenience products, which have more balanced nutritional profiles but less favorable environmental impacts.
By linking geographic context to retail inventory, we provide evidence that reducing food waste requires interventions tailored to local retail dynamics, highlighting the importance of region-specific sustainability strategies.
}
\maketitle

\section*{Introduction}
The growing field of food informatics -- which draws on large-scale digital records to chart the complex system of human interactions with food products -- has transformed our understanding of food processing, consumption, and diet \cite{tufano_data-driven_2025, tilman_global_2014, ispirova_informatics_2025, kuhl_ai_2025}.
For example, advances in computational approaches now allow us to predict the levels of food processing \cite{menichetti_machine_2023}, revealing that more than 73\% of the U.S. food supply is ultra-processed \cite{ravandi_prevalence_2025}, and uncovering the vast, unmapped chemical diversity of our diets \cite{barabasi_unmapped_2019}.
At the same time, integrating mobility data with food choice patterns \cite{horn_population_2023,xu_integrating_2023} has shown that the food environments people encounter throughout their daily routines differ markedly from those near their homes, significantly shaping dietary decisions and consumption behaviors \cite{garcia_bulle_bueno_effect_2024}.
Yet, while this line of research has clarified the link between food and health through nutrition \cite{wingrove_outcomes_2025, lane_ultra-processed_2024, muncke_health_2025, fanzo_sustainable_2022}, food systems also influence human health through another key pathway: food waste, which generates environmental burdens that exacerbate resource scarcity and associated health risks \cite{minor_rising_2022, romanello_2024_2024, nijdam_price_2012, springmann_analysis_2016, wang_food_2025, anshassi_improving_2025, shepon_opportunity_2018}.
Globally, nearly one-third of all food produced for human consumption is lost or discarded each year, with profound environmental, economic, and social consequences \cite{macdiarmid_food_2022, ambikapathi_global_2022, schneider_state_2023, crippa_food_2021}.

While food waste originates from inefficiencies throughout the supply chain \cite{nerin_food_2016, huang_food_2021, campi_specialization_2021, filimonau_exploratory_2017} -- from production through distribution to consumption --  the retail sector represents a critical transition point where products shift from distribution networks to consumers \cite{principato_food_2018,hubner_demand_2013, tian_extracting_2021}.
As a complex adaptive system \cite{fox_consumer_2004, pennacchioli_retail_2014}, retail sales are shaped by both promotional interventions on perishable items and evolving consumer preferences \cite{principato_food_2018, hubner_demand_2013}, creating heterogeneous spatial and temporal patterns in product distribution across stores \cite{moragues-faus_urban_2021}.
For instance, large-scale analyses of grocery purchasing records have linked dietary behavior to health outcomes, revealing spatially structured differences within cities \cite{aiello_large-scale_2019}, while lifestyle-based consumption behaviors form distinct clusters across urban environments \cite{lee_we_2025}, indicating that food choices -- and by extension food waste -- are deeply embedded in cities’ social and spatial fabric \cite{dong_social_2018,seto_hidden_2016}.
Given extensive research demonstrating how hierarchical spatial organization \cite{arcaute_cities_2016} fundamentally shapes urban mobility \cite{lenormand_comparing_2015,bassolas_hierarchical_2019}, economic activity \cite{di_clemente_urbanization_2021}, and overall human behavior patterns \cite{lengyel_role_2020,yang_identifying_2023,pan_urban_2013,bettencourt_urban_2010,sapienza_exposure_2023}, we were inspired to question whether or not similar spatial hierarchies also structure the spatial distribution of promotions of perishable foods -- in other words, how food waste dynamics and promotional strategies vary across different spatial scales and geographic contexts \cite{seto_hidden_2016}.

We address these questions by analyzing spatial patterns of promotional strategies for near-expiry products across Denmark. 
Even in a small country like Denmark, the impact of food waste is significant -- contributing to approximately 2.3 million tonnes of $\mathrm{CO_2}$-equivalent greenhouse gas emissions annually \cite{pacecircular_denmark_foodwaste}, and beyond economic loss and resource depletion, addressing it is essential for achieving sustainable development goals related to climate action, resource efficiency, and social equity \cite{otto_social_2020, campi_specialization_2021, moragues-faus_urban_2021, schneider_state_2023, schneeman_national_2020}.
Furthermore, a previous analysis of Danish retail data focusing on alcohol consumption  demonstrated significant consumer heterogeneity in food purchasing patterns \cite{johansen_food_2006}, with distinct lifestyle-based clusters showing different preferences for healthy versus processed foods, highlighting how food patterns could vary across regions with fundamentally different consumer bases.
Products are discounted by retailers as part of a complex optimization strategy that balances supply constraints, demand variability, and expiration pressures to minimize waste while maximizing revenue \cite{buisman_discounting_2019,kayikci_data-driven_2022,atan_displaying_2025,anderson_how_2019}.
When a product is placed on sale repeatedly over time, this temporal pattern signals systematic inefficiencies in demand forecasting models, inventory optimization protocols, and product lifecycle management strategies \cite{anderson_how_2019}.
Consequently, understanding the spatial heterogeneity of promotional patterns across geographic scales offers a diagnostic lens into how place-based variations in logistics, forecasting, and consumer behavior shape food waste trajectories in retail networks.

\section*{Results}
Salling Group, Denmark's largest retailer, has focused on food waste reduction with a goal to cut food waste by 50\% by 2030 \cite{sallinggroup_foodwaste}. 
As part of this strategy, Salling Group provides a public API (Application Programming Interface) that offers real-time information about discounted products approaching their expiration dates, such as meat, dairy products, and prepared meals.
We collected data on discounted products across Netto stores, which operate 542 locations in Denmark and rank among the country’s top discount supermarket chains (See Supplementary Materials, Section 1 and Methods for data description).
Our dataset captures daily information for each product, recording when it is on sale at a discounted price and where. 
This dataset allows us to examine the complex spatial-temporal patterns of food waste across Denmark over our 153-day period (See Methods for further details).
Furthermore, we manually inspect the dataset and remove offers corresponding to high volumes of products ($>50$ units), representing the 95th percentile, as these tend to correspond to large stock quantities on sale for promotional campaigns rather than genuine near-expiry products.
We combine store location data with OpenStreetMap to compute distances between stores, and each product has been mapped through its corresponding Nutri-Score and Environmental impact scores to enable comprehensive sustainability analysis using OpenFoodFacts (See Methods for further details).
\subsection*{Spatial Clustering and Product Composition Patterns in Denmark.}
To investigate product distribution patterns across Denmark, we analyze data spanning the country's five administrative regions: Hovedstaden (Capital Region), Sjælland (Zealand), Syddanmark (Southern Denmark), Midtjylland (Central Jutland), and Nordjylland (North Jutland) (See Figure \ref{fig_1} \textbf{a}). 
First, we employ hierarchical spatial clustering of store locations using shortest-path distances on the street network to identify densely connected urban cores and their peripheral hinterlands \cite{hutchison_density-based_2013, arcaute_cities_2016}.
We identify groups of stores that are densely connected in terms of street-network travel distance, while leaving more isolated locations unclustered. 
This approach allows us to distinguish compact urban cores from their (diffuse) peripheries (see Methods for full details).
Figure \ref{fig_1} \textbf{b} shows a distinct core–periphery structure \cite{marin_scalar_2024}, as it is characterized by high silhouette scores (0.65), suggesting well-separated and coherent clusters of stores. 
The maximum persistence (0.36), which quantifies the robustness of clusters across multiple resolutions, is low-to-moderate, reflecting that while the Capital Region forms a stable dominant core, peripheral areas display fragmented, low-persistence clustering that aligns with their scattered population distributions.

To quantify both the magnitude of store-product promotional dynamics -- reflecting inventory management decisions and consumer behavior patterns -- and the temporal persistence of discounting activities that signal systematic surplus conditions, we aggregate products by category and compute cumulative daily promotion frequencies across the observation period.
Figure \ref{fig_1} \textbf{c} shows that Cheese, Pork, Ready-to-Eat Meals, Breads, and Yoghurt emerge as the most commonly offered items nationwide.
These categories represent consumer staples with varying shelf lives, ranging from perishable items such as Pork and Bread, which typically remain viable for approximately one week, to longer-lasting products like Cheese and Yoghurt.
Although overall patterns of on sale items are similar across regions, normalized entropy -- a measure of diversity where higher values indicate a more uniform distribution of promotional activity across product categories -- highlights regional distinctions: the Capital Region (Hovedstaden) -- comprising the urban core of Copenhagen and Frederiksberg alongside adjacent rural municipalities such as Fredensborg, Hillerød, and Gribskov -- have a low mean entropy (0.79) and greatest variability across stores within the region.
This means that, on average, fewer products tend to be on sale, and the high variability suggests substantial heterogeneity in promotional strategies within the region.
On the other end of the scale Syddanmark (0.820), Midtjylland (0.818), Nordjylland (0.815), and Sjælland (0.811) display higher and more stable diversity.
These patterns suggest regional differences with respect to on sale product diversity \cite{fox_consumer_2004,buisman_discounting_2019,kayikci_data-driven_2022,atan_displaying_2025,anderson_how_2019}.

\subsection*{Scale-Dependent Retail Product Similarity}
\begin{figure*}[!ht]
    \centering\includegraphics[width=\textwidth]{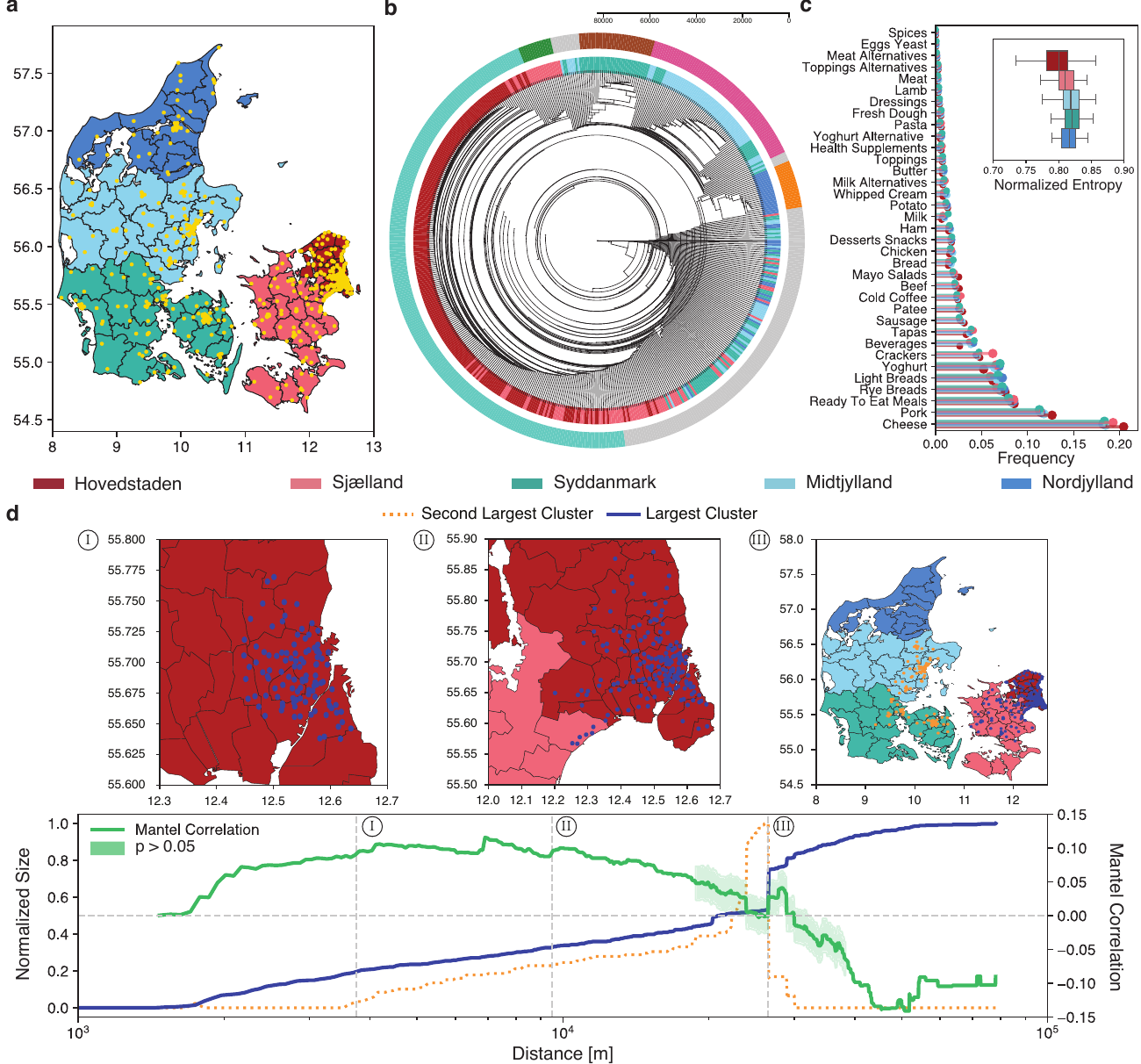}
    \caption{\textbf{Spatial patterns of retail product on sale distribution across Denmark's administrative regions.}
    \textbf{a}) Map of Denmark showing the five administrative regions analyzed: Hovedstaden (red), Sjælland (pink), Syddanmark (teal), Midtjylland (light blue), and Nordjylland (blue), with store locations and spatial clustering results overlaid. Map data copyrighted OpenStreetMap contributors and available from
    \url{https://www.openstreetmap.org}.
    \textbf{b}) Hierarchical spatial clustering analysis showing the core-periphery structure of retail locations based on shortest-path distances along the street network. 
    Clusters are identified by the colors in the second outer ring, with noise shown in grey, whereas regions are identified by the colors in the first ring.
    \textbf{c}) Regional analysis of product-on sale category frequencies and diversity patterns. The lollipop chart, corresponding to a rank plot, shows commonly on sale product categories nationwide (Cheese, Pork, Ready-to-Eat Meals, Breads, and Yoghurt), while the inset of normalized entropy highlights regional differences in product diversity.
    \textbf{d}) Mantel correlation analysis (green) examines the relationship between spatial proximity and promotional composition similarity across multiple distance scales. The correlogram reveals three distinct phases (grey vertical lines): \textbf{(I)} positive spatial-product correlations at smaller scales peaking around $4000$ km, driven by the cohesive Hovedstaden cluster; \textbf{(II)} correlation diminishment upon inclusion of the Sjælland region, with the inversion of the Mantel correlation (and becomes statistically non-significant as highlighted by the shaded area) occurring at the point where the second-largest cluster (orange) merges with the first, as tracked by the blue line (fraction of nodes in largest cluster divided by the maximum value); and \textbf{(III)} negative correlations at larger scales ($\sim26500$ km) indicating compositional divergence among distant stores.    
    }
    \label{fig_1}
\end{figure*}

Building upon these entropy-based findings, we characterize inter-store differences in promotional composition by computing the Bray-Curtis \cite{bray_ordination_1957} dissimilarity matrices across all store pairs, using category-specific promotion frequencies as abundance proxies that reflect underlying supply-demand dynamics (See Methods and Supplementary Materials, Section 2).
The Bray-Curtis dissimilarity index quantifies compositional differences between stores by comparing the relative frequencies of promotional activities across product categories, in direct analogy to ecological analyses that assess community dissimilarity based on species abundance distributions. 
Product categories with high promotional frequencies -- indicative of persistent surplus conditions or systematic demand forecasting errors -- contribute disproportionately to inter-store dissimilarity, with the index ranging from zero (identical promotional profiles) to one (completely distinct promotional patterns), thereby allowing us to understand the spatial heterogeneity of supply chain inefficiencies across retail networks.
Then, we examine the relationship between geographic proximity and compositional similarity using Mantel correlations \cite{diniz-filho_mantel_2013} computed across increasing distance thresholds (see Methods). 
These correlations are summarized in a cumulative correlogram that characterizes how promotional composition similarity varies with spatial separation (Figure~\ref{fig_1}\textbf{d}).

We observe a positive spatial-product correlations at smaller scales, peaking at around $4000$ m (Mantel Correlation $ \sim 0.1$, $p < 0.05$), driven primarily by the cohesive Capital Region cluster (See Figure \ref{fig_1} \textbf{d} \textbf{I}).
This correlation diminishes upon inclusion of the Sjælland region (See Figure \ref{fig_1} \textbf{d} \textbf{II}) and becomes zero and nonsignificant ($p > 0.05$) when peripheral regional clusters merge with the largest connected component ($\sim 26500$ m).
Beyond this critical distance, the correlation starts to be statistically significant, but in the negative direction, indicating divergence in composition among more distant stores (See Figure \ref{fig_1} \textbf{d} \textbf{III}).
This scale-dependent pattern suggests that while local retail environments in the capital region exhibit modest product homogeneity, regional differentiation strategies emerge at larger spatial scales, indicating hierarchical organization in Denmark's retail landscape \cite{arcaute_cities_2016}.
This hierarchical organization is statistically confirmed by PERMANOVA analysis ($F = 6.9$, $p = 0.001$), which tests for differences in multivariate composition using permutation-based comparisons of Bray–Curtis dissimilarities, demonstrating that spatial clusters correspond to genuinely distinct product compositions (see Methods for further details).
To assess the robustness of our findings, we perform a sensitivity analysis by excluding random days from the dataset, incorporating an alternative check that measures total product availability as the sum of individual volumes offered, and normalizing volumes by dividing it by the corresponding population for each store within its respective municipality (See Methods and Supplementary Materials, Section 3). The analysis is unchanged in the light of these perturbations.
\subsection*{Statistical Validated Projection Reveals Regional Retail Communities and Key Products.}
\begin{figure*}[!ht]
    \centering\includegraphics[width=\textwidth]{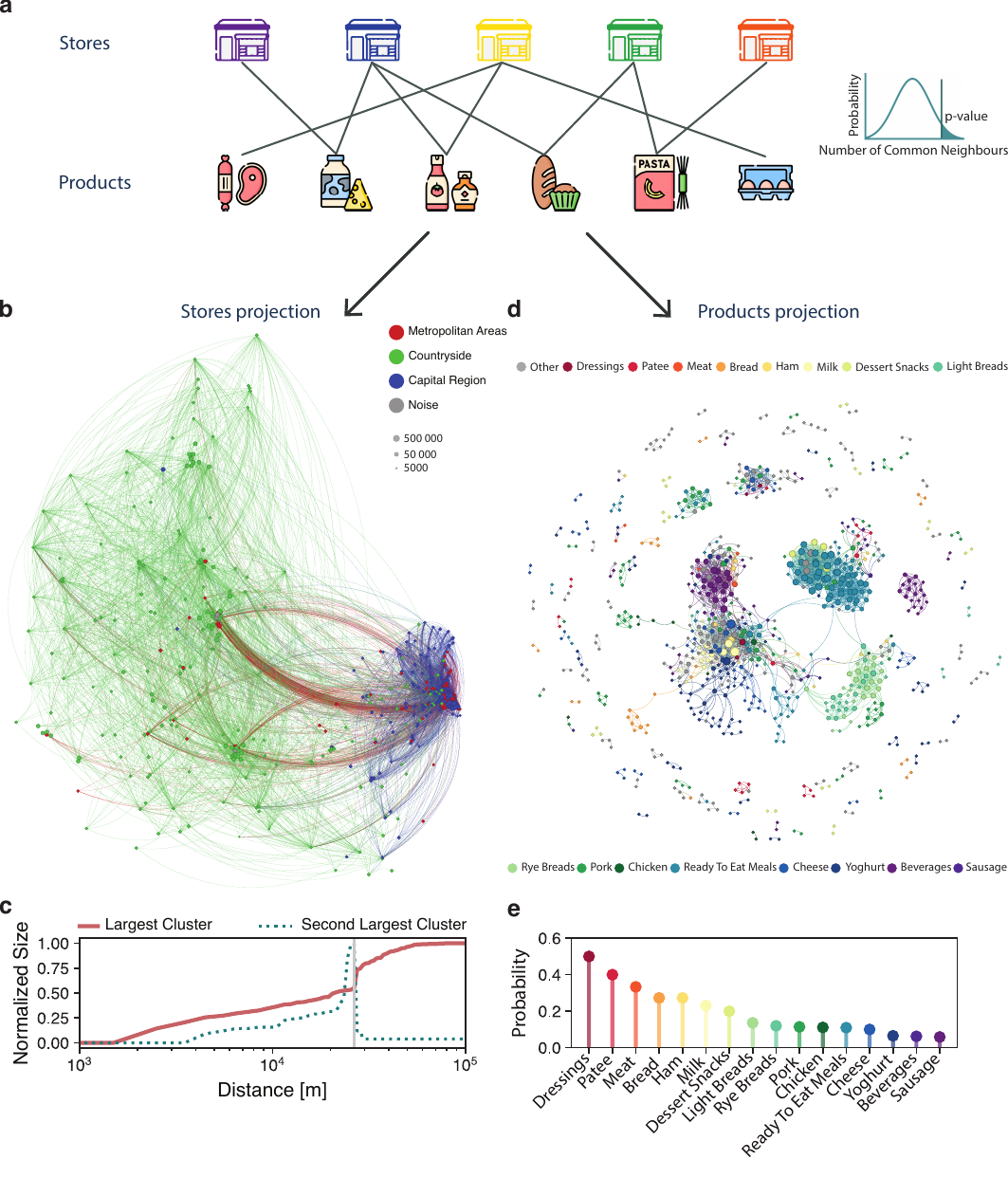}
    \caption{\textbf{Bipartite network analysis identifies product-driven regional retail differentiation in Denmark.}
    \textbf{a}) The infographic summarizes the bipartite network with stores and products on separate layers.
    \textbf{b}) Store-layer projection after validation showing three well-defined communities corresponding to Metropolitan Areas (red), Capital Region (blue), and Countryside (green) clusters. The nodes are positioned according to their geographic coordinates (latitude and longitude), while avoiding overlaps. Node size proportional to the logarithm of the population.
    \textbf{c}) Fraction of nodes in the largest cluster (red) and the second-largest cluster (teal) identified via percolation analysis, with clusters merging at the same distance indicated by the hierarchical clustering (grey vertical line $\sim 26500$ km).    
    \textbf{d}) Products-layer projection using the Yifan Hu layout. Node colors highlight different categories and size proportional to the number of neighbors. The network contains 124 communities, most of which are actually disconnected components (111).
    \textbf{e}) Rank‐ordered probability of each product category appearing among the top 10\% of nodes by betweenness centrality.
    }
    \label{fig_2}
\end{figure*}

Building on the fact that the spatial clustering results reveal scale-dependent patterns in Denmark's retail geography, we use a bipartite network approach to identify which products drive the regional differences (See Figure \ref{fig_2}\textbf{a}).
The bipartite network framework avoids circular reasoning by allowing product–store associations to emerge organically from data rather than relying on geographical distance, which risks conflating spatial proximity with functional relationships. 
The raw network is dense -- comprising 542 stores, 1,854 products, and over 418,000 weighted links, where each edge weight represents the total number of days a specific product appears on sale in a given store -- many of which connections may be redundant or statistically uninformative.
To prune the network we apply two filtering strategies: Revealed Comparative Advantage (RCA) and then the Bipartite Configuration Model (BiCM) to extract the underlying structure from the network \cite{saracco_randomizing_2015, saracco_inferring_2017,cimini_statistical_2019}.
RCA measures whether a store offers products more frequently than expected given the overall market \cite{balassa_revealed_1989}, with values above $1$ indicating relative specialization (See Methods and Supplementary Materials, Section 4 for the RCA distribution).
After removing weaker connections through RCA filtering, we employ BiCM to statistically validate the observed network structure. 
In a nutshell, BiCM generates an ensemble of randomized bipartite networks that preserve the degree sequences of stores (total number of unique products connected) and products (total number of unique stores connected) while randomizing other structural features (See Methods for further details). 
This approach constitutes a null model used to distinguish genuine patterns from artifacts of store size or product popularity (defined here as the number of stores in which a product is on sale) \cite{cimini_statistical_2019}.

When projecting the validated bipartite structure onto the store layer, we identify three well-defined communities (modularity $Q = 0.49$, indicating strong internal connectivity within clusters compared to random connections). 
These clusters correspond closely to the three phases previously revealed as part of our purely spatial analysis: Metropolitan Areas, Capital Region, and Countryside (See Figure \ref{fig_2}\textbf{b}). 
Notably, the Metropolitan Areas cluster extends beyond Copenhagen, encompassing cities such as Odense and Aarhus \cite{marin_scalar_2024}, while the Capital Region cluster reflects the Finger Plan \cite{knowles_transit_2012}, with urban development radiating from central Copenhagen along five suburban ‘fingers’ separated by green wedges.
These communities are robust and emerge from two different community detection methods, Louvain and Infomap \cite{blondel_fast_2008, rosvall_maps_2008} (Adjusted Rand Index $ARI = 0.82$, $p < 0.001$, See Methods for further details).
We incorporate distance weights into the store–store network and apply percolation analysis \cite{arcaute_cities_2016} to identify the hierarchical connectivity structure of this network. 
This percolation profile captures how densely connected store clusters emerge as we vary the distance threshold parameter.
Notably, this distance-weighted percolation profile closely mirrors that of our purely spatial percolation analysis, indicating that spatial proximity information is inherently encoded within the product-based similarity network (See Figure \ref{fig_2}\textbf{c}). 
Collectively, the convergence reveals that a) spatial proximity and b) product on sale similarity offer complementary perspectives on the same underlying organizational principle: stores that operate within similar geographical contexts face comparable market demands and operational constraints. 
These, in turn, generate convergent promotional patterns.
The convergence also highlights potential shared inefficiencies that point to a sharp contrast between rural and urban supply-demand dynamics, and reflecting shared supply chain challenges and consumer preference behaviors \cite{fox_consumer_2004,lee_we_2025,seto_hidden_2016}.
For instance, the consistent discounting of meat in rural areas versus convenience foods in the capital reflect distinct local inefficiencies. 
These disparities suggest that a path to sustainable waste reduction can lie in shifting from reactive uniform discounting to proactive, region-specific inventory adjustments \cite{halloran_addressing_2014}.
As in the preceding section, we perform a sensitivity analysis on the bipartite network weights, as detailed in Supplementary Materials, Section 4, which further underscores the robustness of our results.

On the other hand, the product layer projection captures similarities in aggregate promotional patterns across stores, revealing regionally characteristic product communities and identifying critical ``bridge'' products \cite{pennacchioli_retail_2014}.
This projection has a very high modularity ($Q = 0.83$) and again we find agreement between Infomap and Louvain community assignments (Adjusted Rand Index $ARI = 0.68$, $p < 0.001$). 
The contrast between store-layer modularity ($Q = 0.49$) and product-layer modularity ($Q = 0.83$) points to moderate asymmetric organizational forces \cite{saracco_inferring_2017}: while geographical constraints create moderate store clustering, product relationships exhibit much stronger functional cohesion. 
This cohesion implies that promotional co-occurrence patterns -- driven by shared supply chain dynamics and consumer purchasing behaviors -- create more rigid boundaries in product networks than spatial proximity generates in store networks.
Although more than one hundred distinct communities emerge, this (large) number mainly reflects the fragmented structure of the network (See Figure \ref{fig_2} \textbf{d}). 
Many communities align closely with conventional categories -- Light and Rye Breads, Ready-to-Eat Meals, Beverages, and various Meat assortments -- while notable divergences appear, such as a central cluster that unites Yoghurt, Cheese, Milk, Tapas and  Whipped Cream.
Focusing on the largest connected component, which comprises 433 products, we compute node betweenness centrality (bc) to identify core ``hub'' products that are frequently part of shortest paths between items.
Unlike the categories in Figure \ref{fig_1}\textbf{c}, which are ranked by overall promotional volume, here we identify structural hub products that connect diverse promotional clusters. 
These clusters shape both regional uniformity and inter-regional differences through their network connectivity.
Categories such as Dressings, Meat, Pâté, and Breads are most likely to include nodes with high betweenness centrality ($\text{bc} > 0.03$, top 90th percentile; see Figure \ref{fig_2}\textbf{e}), suggesting they serve as key intermediaries.
%
%
\subsection*{Unhealthy Products Dominate Peripheral Communities.}
Building on the identification of store communities and product clusters, we now turn to the question of what differentiates these clusters in practice \cite{saracco_inferring_2017}.
To address this, we evaluate the per-capita on sale frequency, which reflects how frequently product groups (product categories, nutritional profiles, or environmental ratings) go on sale relative to the number of customers each store serves.
The per-capita on sale frequency is defined as the ratio between validated store–product connections within each community $c$ and product category $k$ (or Nutri-score and Environmental Score), $E_{ck}$, and the total population $P_c$ of distinct stores in the community $c$.
We scale this index by a factor of 100,000, so that it reflects the number of on sale items per 100,000 potential customers.
The per-capita on sale frequency quantifies both oversupply conditions (indicating demand forecasting failures and waste risk) and availability patterns that mirror consumer preferences. 
Thus this measure is able to capture the relationship between supply chain strategies and community characteristics.

\begin{figure*}[!ht]
    \centering\includegraphics[width=\textwidth]{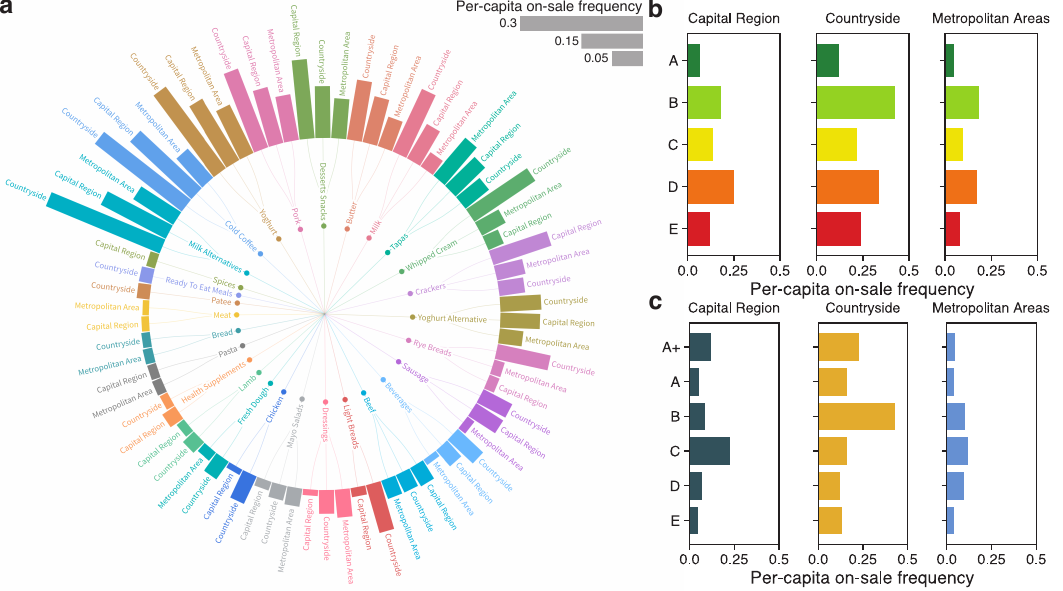}
    \caption{\textbf{Community-specific promotional composition profiles and prevalence of unhealthy products in peripheral danish communities.}
    \textbf{a}) Per-capita on sale frequency (per 100,000 potential customers) for product categories (color) across Metropolitan Areas, Capital Region, and Countryside. The Cheese category has been removed, as its values exceed 0.6 across all communities, and values below 0.01 have also been excluded to improve readability.
    \textbf{b}) Distribution of Nutri-Score ratings (A: healthiest to E: least healthy) across the three community types, showing prevalence of categories B and D and under-representation of the healthiest products in all regions.
    \textbf{c}) Environmental Score distribution (A+: lowest impact to E: highest impact) reveals mid-grade products (B) dominating across communities, with Countryside showing more environmentally positive options while Metropolitan Areas and Capital Region lean toward less environmentally favorable products.
    }
    \label{fig_3}
\end{figure*}

Figure \ref{fig_3} \textbf{a} shows systematic differences in promotional patterns across the three community types, each exhibiting distinct profiles \cite{johansen_food_2006, lee_we_2025}.
A Countryside versus Metropolitan Areas comparison reveals systematic differences in the promotional intensity of meat: Countryside shows a total meat frequency of 0.47 -- 2.2 times higher than the Metropolitan Areas’ 0.21.
In fact Countryside dominates across all meat categories: Pork is nearly twice as frequent as in Metropolitan Areas, while Chicken is absent in the Metropolitan Area.
These patterns suggest that the Countryside has higher promotional intensities for specific meat categories -- Pork, Chicken, Lamb, and Sausage -- than Metropolitan Areas.
Additionally, Butter shows markedly higher promotional intensity in the Countryside, with per-capita on sale frequencies two times greater than those observed in the Metropolitan Areas.
Conversely, products such as Tapas, Pasta, and Crackers are more frequently observed in the Capital Region and the Metropolitan Areas.
Additionally, the Capital Region demonstrates a pronounced preference for convenience and on-the-go products: Cold Coffee, Dessert Snacks, and Crackers all rank among its top five items.
 
To better understand the environmental and nutritional impact of these differences, we have mapped each available product to its corresponding Nutri-Score (72 validated products) and Environmental score (66 validated products), which respectively rate nutritional quality (from A for the healthiest to E for the least healthy) and environmental footprint (from A$+$ for the lowest impact to E for the highest impact).
Figure \ref{fig_3} \textbf{b} shows how categories B and D are the most prevalent across all community types, indicating that the healthiest products (A) are under-represented in promotions \cite{muncke_health_2025}.
Critically, poor nutritional scores (D+E combined) reach 0.58 in Countryside, 0.37 in Capital Region, and 0.25 in Metropolitan Areas, which presents the most balanced profile with a relatively even distribution across all score categories.
In contrast, rural areas exhibit both higher promotional frequencies for high-quality foods and elevated discount rates for low-quality options, suggesting a wider but potentially riskier dietary range \cite{ravandi_prevalence_2025,johansen_food_2006}.
On the other hand, Figure \ref{fig_3} \textbf{c} suggests a complementary picture regarding the Environmental score, where mid-grade (B) dominates overall, and the Countryside demonstrates environmentally positive products.
However, meat products are not mapped, despite being well known to exert a disproportionately large environmental impact \cite{springmann_analysis_2016}.
Notably, the Metropolitan Areas and Capital Region have a distribution shifted towards less environmentally positive products (C-D).
%
%
\section*{Discussion}

Human interactions with food systems generate system-wide challenges \cite{schneider_state_2023}, creating complex feedback loops between dietary behaviors, environmental impacts, and social outcomes that reinforce one another \cite{filimonau_exploratory_2017,huang_food_2021,wang_food_2025,anshassi_improving_2025,macdiarmid_food_2022}.
Our analysis reveals regional differences in the near-expiry discounted products across Denmark, demonstrating that spatial context shapes both the composition and intensity of near-expiry product offerings \cite{lengyel_role_2020}.
The bipartite approach identifies three distinct retail communities -- Countryside, Metropolitan Areas, and Capital Region -- each exhibiting unique profiles in meat, dairy, convenience, nutritional, and environmental product assortments \cite{saracco_inferring_2017,cimini_statistical_2019}.
Our findings highlight the need for food waste reduction strategies that are sensitive to local retail dynamics \cite{seto_hidden_2016}, as neglecting such spatial differences may reinforce supply chain inefficiencies and amplify environmental burdens.
While general interventions, such as the phasing out of quantity discounts, have been effective in Denmark, our results suggest that further reductions could arise from regionally differentiated policies -- for example, targeting single-household packaging sizes in capital regions to reduce convenience food waste, while focusing on cold-chain inventory optimization for meat products in rural areas \cite{halloran_addressing_2014}.

Our work comes with limitations.
First, our study relies on 153 days of data from a single retail chain in Denmark, which may not capture the full spectrum of retail behaviors across all supermarket formats.
Recognizing the temporal limitation, we have performed the same analysis removing days from the dataset to ensure the robustness of our main results (See Supplementary Materials, Section 3).
Second, our approach relies on promotional frequency. 
As discussed above, this is a composite proxy that integrates multiple underlying drivers, including demand forecasting accuracy, inventory management efficiency, supply chain dynamics, and consumer behavioral patterns -- rather than direct measures of spoilage or waste quantities, limiting our ability to disentangle the precise contribution of each factor.
Third, the mapping of products to Nutri-Score and Environmental scores covers only a subset of items (see Methods), potentially introducing biases in the assessment of nutritional and environmental profiles across regions.

Overall, our results underscore the strong role of geography in shaping human behavior patterns \cite{arcaute_cities_2016, dong_social_2018, lengyel_role_2020, sapienza_exposure_2023}.
Food choices are grounded in local social and spatial contexts, influenced by the organization of urban and rural areas and how promotions are tailored to these settings.
The regional differences in per-capita on sale frequencies and product composition patterns highlight that supply-demand mismatches vary across locations, emphasizing that effective food waste reduction strategies must be tailored to local retail dynamics rather than applied uniformly \cite{dogan_cities_2023}.
Retail patterns shift across regions and lifestyle clusters, showing how deeply place and daily decisions are linked \cite{pan_urban_2013}.
Food waste, then, is the sum of these everyday choices -- highlighting the need for strategies that tackle both geographic disparities and individual behavior to drive sustainable consumption \cite{otto_social_2020}.
\section*{Methods}
\subsection*{Data Descriptions}

We have retrieved Data from the Salling Group Anti Food Waste API, encompassing both store and product information (See Supplementary Materials, Section 1 for all the details). 
Store data include geographic coordinates, unique identifiers for each location and store brand.
Product data capture items on discount offers, including EAN codes, product categories, stock levels, and temporal information such as discount start and end times. 
Additionally, we have mapped each product to its corresponding Nutri-Score and environmental score using OpenFoodFacts, an open database of food products from around the world that aggregates ingredient lists, nutritional information, and sustainability metrics. 
Nutri-Score is a front-of-pack labeling system that grades foods from A (healthiest) to E (least healthy) based on their nutritional composition, while the environmental score quantifies the product’s ecological footprint across factors like greenhouse gas emissions, water usage, and land use \cite{andreani_nutri-score_2025}. 
We have collected Data during three predefined periods (2024-08-08 to 2024-10-10, 2025-03-05 to 2025-04-10, and 2025-05-01 to 2025-06-21) and analyses are limited to these observation windows.
When multiple discount offers existed for the same product on the same day at a given store, only the record with the lowest price has been retained. 
This processing approach results in a structured dataset that captures the daily availability of discounted products across the Netto store network throughout 153 days.
We have computed the shortest road‐network distance between every pair of stores using the Open Source Routing Machine (OSRM) on OpenStreetMap data \cite{huber_osrmtime_2016}, and obtained municipality boundaries and resident population figures for Copenhagen’s level‐2 administrative units (municipality level) from the Copernicus Global Degree of Urbanisation Classification dataset \cite{european_commission_joint_research_centre_ghsl_2023}.
\subsection*{Hierarchical Clustering}

To identify spatially coherent retail clusters, we have employed Hierarchical Density-Based Spatial Clustering of Applications with Noise (HDBSCAN), a density-based clustering algorithm that extends DBSCAN by converting it into a hierarchical clustering framework \cite{hutchison_density-based_2013}. 
HDBSCAN constructs a hierarchy of clusters by varying the density parameter and extracting stable clusters across multiple density levels. 
We have used the pre-computed shortest path distance matrix between all store locations and set the minimum cluster size parameter to 10 stores to ensure that identified clusters represent meaningful retail agglomerations rather than isolated store pairs.
The clustering procedure identified five distinct spatial clusters across Denmark's retail landscape, grouping part of Sjælland and all of Hovedstaden into one cluster, combining a mix of Midtjylland and part of Syddanmark in another, capturing a separate part of Syddanmark in a third, forming clusters corresponding mainly to Nordjylland and Sjælland, while noise is scattered primarily across Jutland.
To assess clustering quality, we have computed three complementary metrics: the Davies-Bouldin Index (0.44), which measures the average similarity ratio between clusters where lower values indicate better separation; the Silhouette Score (0.65), quantifying how well-separated clusters are with values approaching 1.0 indicating optimal clustering; and Maximum Persistence (0.364), representing the stability of clusters across the hierarchical structure where lower values suggest less robust cluster boundaries.

To quantify promotional composition differences between stores, we have computed Bray-Curtis dissimilarity indices for all store pairs based on their category-specific promotion frequencies. The Bray-Curtis index \cite{bray_ordination_1957} is defined as
\begin{equation}
\text{Bray--Curtis Similarity}_{ij} = 1 - \frac{\sum_{k=1}^{n} |x_{ik} - x_{jk}|}{\sum_{k=1}^{n} (x_{ik} + x_{jk})}\,,
\end{equation}
where $x_{ik}$ and $x_{jk}$ represent the abundance of product category $k$ in stores $i$ and $j$, respectively, and $n$ is the total number of categories (See Supplementary Materials, Section 2 for the dendrogram).
Abundance for each product category is calculated as the cumulative number of products on sale multiplied by the number of days they were offered.
We have assessed the relationship between spatial proximity and product similarity using Mantel correlation analysis across multiple distance scales. 
The Mantel test evaluates whether two distance matrices are significantly correlated by computing Pearson correlations between corresponding elements and testing significance through permutation \cite{diniz-filho_mantel_2013}.
We have constructed a cumulative correlogram by calculating Mantel correlations at increasing distance thresholds, enabling detection of scale-dependent spatial autocorrelation patterns in promotional composition similarity.
To statistically validate that identified spatial clusters correspond to genuinely distinct product composition strategies, we have performed Permutational Multivariate Analysis of Variance (PERMANOVA) using the Bray-Curtis dissimilarity matrix and the spatial clusters. 
PERMANOVA is a non-parametric test that evaluates whether groups differ in multivariate composition by partitioning distance matrices and calculating pseudo F-statistics. 
Group labels are permuted 1000 times to generate a null F-distribution, and significance is assessed by comparing the observed F to this permutation-based distribution, avoiding assumptions about data distribution.
\subsection*{Bipartite Configuration Model and Network Analysis}

To identify products driving regional retail differentiation, we have built a bipartite network linking stores $s$ to products $p$ based on the cumulative number of days each product was on sale, $X_{s,p}$.
First, we have applied RCA filtering to identify meaningful store-product associations.  
RCA evaluates whether a store puts on offer certain products more often than would be expected based on overall market trends \cite{balassa_revealed_1989}. It is calculated as the ratio of a store's share of a particular product to the product's share of total offerings across all stores:
\begin{equation}
\text{RCA}_{s,p} = \frac{\dfrac{X_{s,p}}{\sum\limits_{p'} X_{s,p'}}}{\dfrac{\sum\limits_{s'} X_{s',p}}{\sum\limits_{s',p'} X_{s',p'}}}\,,
\end{equation}
where $\sum_{p'} X_{s,p'}$ is total number of product-days in store $s$, $\sum_{s'} X_{s',p}$ is the total number of days product $p$ was on sale across all stores, and $\sum_{s',p'} X_{s',p'}$ total number of product-days across all stores and all products. RCA values greater than 1 indicate relative specialization (See Supplementary Materials, Section 4 for the RCA distribution).

The RCA filtering yields an unweighted bipartite network that can be validated via the Bipartite Configuration Model \cite{saracco_randomizing_2015,saracco_inferring_2017,cimini_statistical_2019}.
BiCM defines a maximum-entropy null ensemble constrained by the observed degree sequences -- that is, the number of unique products connected to each store and vice versa. 
The model assigns connection probabilities
\begin{equation}
p_{sc} = \frac{e^{-\lambda_s -\theta_c}}{1+e^{-\lambda_s -\theta_c}}\,,
\end{equation}
where $\{\lambda_s\}$ and $\{\theta_c\}$ are the multipliers enforcing the degree constraints for stores and products, respectively.
From these probabilities, the expected number of common stores between two products $a$ and $b$ is just $\langle V_{ab} \rangle = \sum_s p_{sa} p_{sb}$, and the observed co‐occurrence $V_{ab}$ is tested against its Poisson–Binomial distribution via $p[V_{ab}] = 1-\sum_{x = 0}^{V_{ab} -1} \pi (x\vert a,b)$.
Finally, we apply False Discovery Rate (FDR) to control for multiple hypothesis testing when validating individual links (significance level $\alpha  = 0.05$), ensuring that retained connections represent statistically significant deviations from random expectation rather than spurious correlations.

We perform community detection on the resulting projections using Infomap, which identifies modules by modeling information flow on the network as a random walk and optimizing a map equation \cite{rosvall_maps_2008}, and Louvain, which greedily maximizes modularity \cite{blondel_fast_2008} -- a measure of the density of edges inside communities versus edges between communities.
The two resulting partitions were then compared using the Adjusted Rand Index (ARI) to assess their similarity based on the contingency matrix and statistical significance via permutation testing.

Percolation on the weighted (store-store) network is performed by using travel distance as the link weight. At a given distance threshold, only edges shorter than that threshold are retained (equivalently, distances can be normalized to bond-open probabilities), allowing connected stores to form clusters that merge as the threshold increases \cite{arcaute_cities_2016}.
\section*{Declarations}
\begin{itemize}
\item Data Availability:
The Salling Group API provides open access to food waste data through their developer portal \url{https://developer.sallinggroup.com/api-reference}.
OSRM provides publicly accessible routing services through their API endpoints at \url{http://router.project-osrm.org/}.
The Nutri-Score and environmental score mappings were obtained by querying the Open Food Facts database via its REST API \url{https://openfoodfacts.github.io/openfoodfacts-server/api/}.
Municipality boundaries and resident population were obtained from the Copernicus Global Degree of Urbanisation Classification dataset \url{https://human-settlement.emergency.copernicus.eu/ghs_duc2023.php}.
\item Code Availability:
The data and code to reproduce the analysis is released on \href{https://github.com/RiegelGestr/FoodWaste}{GitHub}.
For inquiries, please contact A.D. \url{antde@dtu.dk}.
\item Acknowledgements
A.D. acknowledges the helpful discussions and insightful feedback with Riccardo Di Clemente, Yuan Liao, Giovanni Mauro, Fabio Saracco and Michele Tizzani.
\item Author Contributions:
A.D. and F.B. gathered the data.
F.B. performed preliminary analysis.
A.D. and A.G. performed the analysis and made the figures.
A.D. and S.L. designed the analysis.
S.L. supervised the project.
All authors discussed the results and contributed to the final manuscript.
\item Competing Interests:
The authors declare no competing interests.
\item Funding Declaration:
Funding not applicable.
\end{itemize}
\bibliography{bibliography.bib}
\pagebreak[2]
\clearpage
\renewcommand{\figurename}{Supplementary Figure}
\renewcommand{\tablename}{Supplementary Table}
\setcounter{figure}{0}
\renewcommand{\thefigure}{S\arabic{figure}}
\section*{Supplementary Materials, Section 1: Dataset Information}\label{sm_dataset_info}

\begin{table}[h!]
	\centering
	\begin{tabular}{lp{0.6\textwidth}}
		\toprule
		{\textbf{Column}} & {\textbf{Description}}\\
		\midrule
		Store ID &  ID that uniquely identifies each store\\
		Position & Latitude and longitude position of each store\\
		Store Brand & Store Brand\\
		\hline 
	\end{tabular}
    \caption{Metadata downloaded from Salling Group API for each store.
    } 
\label{tab_sm_data_store}
\end{table}
\begin{table}[h!]
	\centering
	\begin{tabular}{lp{0.6\textwidth}}
		\toprule
		{\textbf{Column}} & {\textbf{Description}}\\
		\midrule
        Product Name   & Name of the product on offer \\
        Product ID     & EAN code of the product \\
        Category       & Category of the product as provided by the data provider \\
        Start Day      & Start date when the product is on offer \\
        End Day        & End date when the product is on offer \\
        Current Day    & Current date of data retrieval \\
        Stock          & Quantity of the product available in store \\
        Store ID       & Unique identifier of the store where the product is on offer \\
        \hline 
	\end{tabular}
    \caption{Metadata downloaded from Salling Group API for each product on offer.
    } 
\label{tab_sm_data_product}
\end{table}
\begin{table}[h!]
	\centering
	\begin{tabular}{lp{0.6\textwidth}}
		\toprule
		{\textbf{Column}} & {\textbf{Description}}\\
		\midrule
        Product ID     & EAN code of the product \\
        Nutriscore       & Nutriscore \\
        Environmental Score      & Environmental Score \\
        \hline 
	\end{tabular}
    \caption{Metadata downloaded from OpenFoodFacts.
    } 
\label{tab_sm_data_product_openfoodfacts}
\end{table}

\begin{table}[h!]
\resizebox{\textwidth}{!}{
\begin{tabular}{ll}
\toprule
Original Category & Mapped Category \\
\midrule
Beverages/Energy Drinks Sport Drinks/Sport Drinks & beverages \\
Beverages/Juice Smoothies/Apple Juices & beverages \\
Beverages/Juice Smoothies/Fruit Shots Vegetable Shots & beverages \\
Beverages/Juice Smoothies/Multi Fruit Juices & beverages \\
Beverages/Juice Smoothies/Orange Juices & beverages \\
Beverages/Juice Smoothies/Other Fruit Juices & beverages \\
Beverages/Juice Smoothies/Smoothies & beverages \\
Bread And Cakes/Bread Crumbs & bread \\
Bread And Cakes/Bread Specialities/Burger Buns & bread \\
Bread And Cakes/Bread Specialities/Hotdog Buns & bread \\
Bread And Cakes/Bread Specialities/Naan Breads Flat Breads & bread \\
Bread And Cakes/Bread Specialities/Pita Bread & bread \\
Bread And Cakes/Bread Specialities/Pizza Crusts & bread \\
Bread And Cakes/Bread Specialities/Special Whole Breads & bread \\
Bread And Cakes/Bread Specialities/Tartlets & bread \\
Bread And Cakes/Crackers Crisp Bread/Crackers/Chocolate Crackers Sweet Crackers & crackers \\
Bread And Cakes/Crackers Crisp Bread/Crackers/Croutons Breadsticks & crackers \\
Bread And Cakes/Crackers Crisp Bread/Crackers/Rice Crackers Corn Crackers & crackers \\
Bread And Cakes/Crackers Crisp Bread/Crackers/Salted Crackers & crackers \\
Bread And Cakes/Crackers Crisp Bread/Crackers/Whole Grain Crackers & crackers \\
Bread And Cakes/Crackers Crisp Bread/Crisp Bread & crackers \\
Bread And Cakes/Crackers Crisp Bread/Rusks & crackers \\
Bread And Cakes/Light Breads Dark Breads/Light Dark Buns & light breads \\
Bread And Cakes/Light Breads Dark Breads/Toast Breads & light breads \\
Bread And Cakes/Light Breads Dark Breads/Whole Breads & light breads \\
Bread And Cakes/Rye Breads/Other Rye Breads & rye breads \\
Bread And Cakes/Rye Breads/Rye Breads With Carrots & rye breads \\
Bread And Cakes/Rye Breads/Seeded Rye Breads & rye breads \\
Bread And Cakes/Rye Breads/Sun Flower Seeded Rye Breads & rye breads \\
Dairy And Cold Storage/Bacon Sausages Toppings/Bacon/Bacon Whole Pieces & sausage \\
Dairy And Cold Storage/Bacon Sausages Toppings/Bacon/Chopped Bacon Bits & sausage \\
Dairy And Cold Storage/Bacon Sausages Toppings/Bacon/Sliced Bacon & ham \\
Dairy And Cold Storage/Bacon Sausages Toppings/Chopped Ham & ham \\
Dairy And Cold Storage/Bacon Sausages Toppings/Other Toppings & toppings \\
Dairy And Cold Storage/Bacon Sausages Toppings/Sausages/Beef Sausages & sausage \\
Dairy And Cold Storage/Bacon Sausages Toppings/Sausages/Chicken Sausages & chicken \\
Dairy And Cold Storage/Bacon Sausages Toppings/Sausages/Cocktail Sausages Brunch Sausages & sausage \\
Dairy And Cold Storage/Bacon Sausages Toppings/Sausages/Pork Sausages & pork \\
Dairy And Cold Storage/Bacon Sausages Toppings/Sausages/Pork Sausages/Meat Fish/Pork/Other Pork & pork \\
Dairy And Cold Storage/Bacon Sausages Toppings/Sausages/Veggie Sausages & meat alternatives \\
Dairy And Cold Storage/Dairy/Butter Fats/Butter & butter \\
Dairy And Cold Storage/Dairy/Butter Fats/Fats & butter \\
Dairy And Cold Storage/Dairy/Butter Fats/Margarine & butter \\
Dairy And Cold Storage/Dairy/Butter Fats/Other Butters & butter \\
Dairy And Cold Storage/Dairy/Cheese/Blue Cheese Brie & cheese \\
Dairy And Cold Storage/Dairy/Cheese/Cheese For Slicing & cheese \\
Dairy And Cold Storage/Dairy/Cheese/Cheese Specialities & cheese \\
Dairy And Cold Storage/Dairy/Cheese/Cottage Cheese & cheese \\
Dairy And Cold Storage/Dairy/Cheese/Cream Cheese & cheese \\
Dairy And Cold Storage/Dairy/Cheese/Grilled Cheese & cheese \\
Dairy And Cold Storage/Dairy/Cheese/Mozarella & cheese \\
Dairy And Cold Storage/Dairy/Cheese/Salat Cheese & cheese \\
Dairy And Cold Storage/Dairy/Cheese/Shredded Cheese Toppings & cheese \\
Dairy And Cold Storage/Dairy/Cheese/Sliced Cheese & cheese \\
Dairy And Cold Storage/Dairy/Cheese/Snack Cheese & cheese \\
Dairy And Cold Storage/Dairy/Cold Butter Milk Soup & yoghurt \\
Dairy And Cold Storage/Dairy/Cold Coffee Drinks Cacao & cold coffee \\
Dairy And Cold Storage/Dairy/Cream Whipped Cream/Coffee Cream & whipped cream \\
Dairy And Cold Storage/Dairy/Cream Whipped Cream/Cooking Cream & whipped cream \\
Dairy And Cold Storage/Dairy/Cream Whipped Cream/Whipped Cream & whipped cream \\
Dairy And Cold Storage/Dairy/Cream Whipped Cream/Whipping Cream & whipped cream \\
Dairy And Cold Storage/Dairy/Desserts Snacks & desserts snacks \\
Dairy And Cold Storage/Dairy/Desserts Snacks/Dairy And Cold Storage/Ready To Eat Meals/Dinner Meals & ready to eat meals \\
Dairy And Cold Storage/Dairy/Milk/Milk Drinks & milk \\
Dairy And Cold Storage/Dairy/Milk/Nonfat Milk & milk \\
Dairy And Cold Storage/Dairy/Milk/Semiskimmed Milk & milk \\
Dairy And Cold Storage/Dairy/Milk/Skimmed Milk & milk \\
Dairy And Cold Storage/Dairy/Milk/Whole Milk & milk \\
Dairy And Cold Storage/Dairy/Plant Based Alternatives/Plant Based Butter Fat Alternatives & butter \\
Dairy And Cold Storage/Dairy/Plant Based Alternatives/Plant Based Cream Alternatives & yoghurt alternative  \\
Dairy And Cold Storage/Dairy/Plant Based Alternatives/Plant Based Milk Alternatives & milk alternatives \\
Dairy And Cold Storage/Dairy/Plant Based Alternatives/Plant Based Yoghurt Alternatives & yoghurt alternative  \\
Dairy And Cold Storage/Dairy/Yoghurt Soured Milk Products/Buttermilk & yoghurt \\
Dairy And Cold Storage/Dairy/Yoghurt Soured Milk Products/Creme Fraiche Quark & yoghurt \\
Dairy And Cold Storage/Dairy/Yoghurt Soured Milk Products/Drinking Yoghurt Kids Yoghurt & yoghurt \\
Dairy And Cold Storage/Dairy/Yoghurt Soured Milk Products/Skyr & yoghurt \\
Dairy And Cold Storage/Dairy/Yoghurt Soured Milk Products/Soured Milk & yoghurt \\
Dairy And Cold Storage/Dairy/Yoghurt Soured Milk Products/Yoghurt & yoghurt \\
Dairy And Cold Storage/Dairy/Yoghurt Soured Milk Products/Yoghurt/Dairy And Cold Storage/Dairy/Yoghurt Soured Milk Products/Skyr & yoghurt \\
Dairy And Cold Storage/Eggs Yeast/Yeast & eggs yeast \\
Dairy And Cold Storage/Fresh Dough And Pancakes/fresh Dough & fresh dough \\
Dairy And Cold Storage/Fresh Dough And Pancakes/Pancakes & fresh dough \\
Dairy And Cold Storage/Fresh Dough And Pancakes/Prepared Pizza Base & fresh dough \\
\bottomrule
\end{tabular}
\hspace{1cm}
\begin{tabular}{ll}
\toprule
Original Category & Mapped Category \\
\midrule
Dairy And Cold Storage/Fresh Pasta/Fresh Lasagna Layers & pasta \\
Dairy And Cold Storage/Fresh Pasta/Fresh Pasta With Fillings & pasta \\
Dairy And Cold Storage/Fresh Pasta/Fresh Pasta Without Fillings & pasta \\
Dairy And Cold Storage/Lunch Meats/Attachments For Lunch Meats & ready to eat meals \\
Dairy And Cold Storage/Lunch Meats/Liver Paste Patee & patee \\
Dairy And Cold Storage/Lunch Meats/Mayo Salads/Fish Mayo Salads & mayo salads \\
Dairy And Cold Storage/Lunch Meats/Mayo Salads/Italian Mayo Salads & mayo salads \\
Dairy And Cold Storage/Lunch Meats/Mayo Salads/Other Mayo Salads & mayo salads \\
Dairy And Cold Storage/Lunch Meats/Mayo Salads/Pork Mayo Salads & mayo salads \\
Dairy And Cold Storage/Lunch Meats/Mayo Salads/Poultry Mayo Salads & mayo salads \\
Dairy And Cold Storage/Lunch Meats/Meat Balls & meat \\
Dairy And Cold Storage/Lunch Meats/Meat/Cold Cuts Chicken & chicken \\
Dairy And Cold Storage/Lunch Meats/Meat/Cold Cuts Ham & ham \\
Dairy And Cold Storage/Lunch Meats/Meat/Cold Cuts Turkey & chicken \\
Dairy And Cold Storage/Lunch Meats/Meat/Corned Beef & beef \\
Dairy And Cold Storage/Lunch Meats/Meat/Meat Sausage & sausage \\
Dairy And Cold Storage/Lunch Meats/Meat/Roastbeef & beef \\
Dairy And Cold Storage/Lunch Meats/Meat/Salami & pork \\
Dairy And Cold Storage/Lunch Meats/Meat/Salami/Dairy And Cold Storage/Tapas Specialities/Cold Cuts Specialities & tapas \\
Dairy And Cold Storage/Lunch Meats/Meat/Salami/Dairy And Cold Storage/Tapas Specialities/Sausages Specialities & tapas \\
Dairy And Cold Storage/Lunch Meats/Meat/Sausage Roll & sausage \\
Dairy And Cold Storage/Lunch Meats/Meat/Sliced Lunch Meats & ready to eat meals \\
Dairy And Cold Storage/Lunch Meats/Meat/Smoked Saddle Of Pork & pork \\
Dairy And Cold Storage/Lunch Meats/Plant Based Toppings & toppings alternatives \\
Dairy And Cold Storage/Lunch Meats/Prepared Chicken & chicken \\
Dairy And Cold Storage/Potato Accessories/Cream Potatoes & potato \\
Dairy And Cold Storage/Potato Accessories/potato Salad & potato \\
Dairy And Cold Storage/Potato Accessories/Prepared Potatoes & potato \\
Dairy And Cold Storage/Ready To Eat Meals/Dinner Meals & ready to eat meals \\
Dairy And Cold Storage/Ready To Eat Meals/Dinner Meals/Meat Fish/Pork/Other Pork & ready to eat meals \\
Dairy And Cold Storage/Ready To Eat Meals/Dinner Salads & ready to eat meals \\
Dairy And Cold Storage/Ready To Eat Meals/Ready To Eat Meat Alternatives & ready to eat meals \\
Dairy And Cold Storage/Ready To Eat Meals/Ready To Eat Pizza & ready to eat meals \\
Dairy And Cold Storage/Ready To Eat Meals/Ready To Eat Soups & ready to eat meals \\
Dairy And Cold Storage/Ready To Eat Meals/Sandwiches Wraps & ready to eat meals \\
Dairy And Cold Storage/Sauces Dressings/Dressings & dressings \\
Dairy And Cold Storage/Sauces Dressings/Remoulade Mayonnaise & dressings \\
Dairy And Cold Storage/Sauces Dressings/Sauce & dressings \\
Dairy And Cold Storage/Tapas Specialities/Cold Cuts Specialities & tapas \\
Dairy And Cold Storage/Tapas Specialities/Olives Other Attachments & tapas \\
Dairy And Cold Storage/Tapas Specialities/Pesto Hummus & tapas \\
Dairy And Cold Storage/Tapas Specialities/Sausages Specialities & tapas \\
Dairy And Cold Storage/Tapas Specialities/Sausages Specialities/Dairy And Cold Storage/Tapas Specialities/Cold Cuts Specialities & tapas \\
Food Of The World/Asian Food/Asian Sauces/Asian Dressings & dressings \\
Food Of The World/Mexican Food/Mexican Dressing & dressings \\
Food Of The World/Mexican Food/Mexican Dressing/Dairy And Cold Storage/Sauces Dressings/Remoulade Mayonnaise & dressings \\
Food Of The World/Mexican Food/Salsa & dressings \\
Frozen Products/Frozen Meat Fish/Frozen Pork & pork \\
Frozen Products/Ready Meals/Asian Ready Meals & ready to eat meals \\
Frozen Products/Ready Meals/Asian Ready Meals/Dairy And Cold Storage/Ready To Eat Meals/Dinner Meals & ready to eat meals \\
Frozen Products/Ready Meals/Other Frozen Ready Meals & ready to eat meals \\
Grocery/Cooking Oils Vinegar Dressings Sauces/Dressings Attachments/Mayonnaise Aioli & dressings \\
Grocery/Cooking Oils Vinegar Dressings Sauces/Dressings Attachments/Salad Dressing & dressings \\
Grocery/Ready Meals Soups/Other Ready Meals & ready to eat meals \\
Grocery/Tinned Food Broth Spices/Pestos Tapenades/Tapenades & tapas \\
Grocery/Tinned Food Broth Spices/Spices Broth Fond/Spices & spices \\
Meat Fish/Beef/Minced Beef & beef \\
Meat Fish/Beef/Roasted Whole Beef Pieces & beef \\
Meat Fish/Chicken/Chicken Breasts Chicken Filets & chicken \\
Meat Fish/Chicken/Chopped Chicken & chicken \\
Meat Fish/Chicken/Other Chicken & chicken \\
Meat Fish/Chicken/Other Chicken/Dairy And Cold Storage/Ready To Eat Meals/Dinner Meals & ready to eat meals \\
Meat Fish/Lamb & lamb \\
Meat Fish/Meat Alternatives & meat alternatives \\
Meat Fish/Pork/Chopped Pork & pork \\
Meat Fish/Pork/Minced Pork & pork \\
Meat Fish/Pork/Minced Pork Veal & pork \\
Meat Fish/Pork/Other Pork & pork \\
Meat Fish/Pork/Other Pork/Dairy And Cold Storage/Bacon Sausages Toppings/Sausages/Pork Sausages & pork \\
Meat Fish/Pork/Other Pork/Dairy And Cold Storage/Lunch Meats/Liver Paste Patee & patee \\
Meat Fish/Pork/Other Pork/Dairy And Cold Storage/Ready To Eat Meals/Dinner Meals & ready to eat meals \\
Meat Fish/Pork/Other Pork/Frozen Products/Frozen Meat Fish/Frozen Pork & pork \\
Meat Fish/Pork/Other Pork/Meat Fish/Pork/Roasted Whole Pork Pieces/Roasted Pork & pork \\
Meat Fish/Pork/Pork Chops Pork Schnitzels & pork \\
Meat Fish/Pork/Roasted Whole Pork Pieces/Ham & ham \\
Meat Fish/Pork/Roasted Whole Pork Pieces/Pork Neck Pork Loin & pork \\
Meat Fish/Pork/Roasted Whole Pork Pieces/Pork Neck Pork Loin/Meat Fish/Pork/Other Pork & pork \\
Meat Fish/Pork/Roasted Whole Pork Pieces/Pork Tenderloin & pork \\
Meat Fish/Pork/Roasted Whole Pork Pieces/Roasted Pork & pork \\
Meat Fish/Pork/Roasted Whole Pork Pieces/Roasted Pork/Meat Fish/Pork/Roasted Whole Pork Pieces/Pork Neck Pork Loin & pork \\
Meat Fish/Pork/Thick Sausages & pork \\
Meat Fish/spareribs Barbecue Meat & beef \\
Personal Care/Health Supplements/Protein Shakes Protein Bars/Protein Shake & health supplements \\
\bottomrule
\end{tabular}
}
    \caption{Category from data provider and corresponding mapped category.
    } 
\label{tab_sm_data_product_cat}
\end{table}
\pagebreak[2]
\clearpage
\section*{Supplementary Materials, Section 2: Bray Curtis Similarity between Stores}\label{sm_bc}
\begin{figure*}[!ht]
    \centering\includegraphics[width=\textwidth]{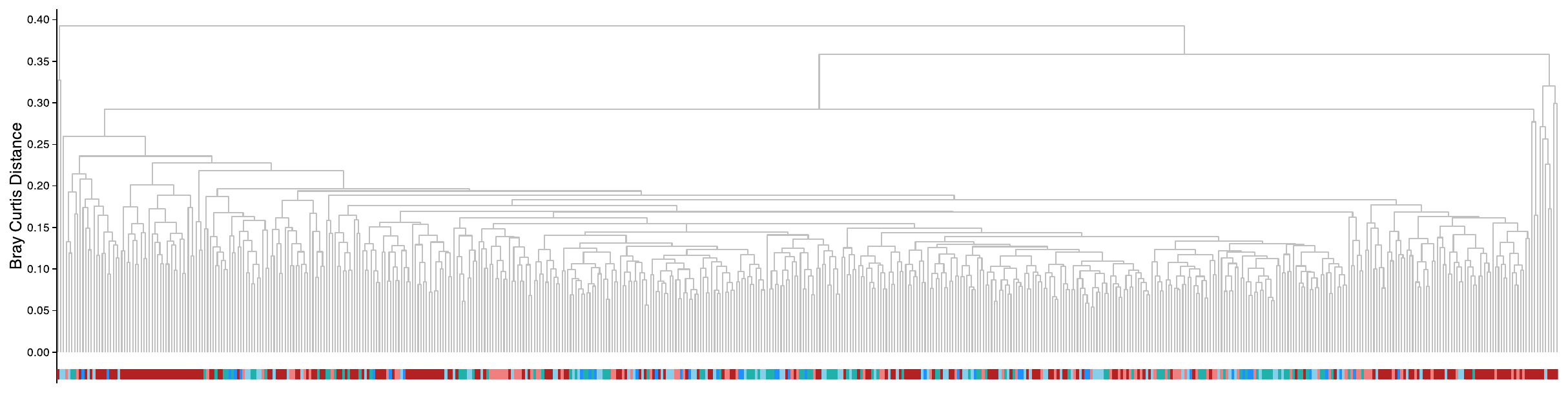}
    \caption{\textbf{Dendrograms of stores based on Bray–Curtis similarity.}
    Each leaf represents a store, with adjacent colored bars denoting the corresponding region: Hovedstaden (red), Sjælland (pink), Syddanmark (teal), Midtjylland (light blue), and Nordjylland (blue).
    We observe a tendency for stores belonging to the same region to cluster more closely within the dendrogram.
    }
    \label{fig_sm_bc}
\end{figure*}
\pagebreak[2]
\clearpage
\section*{Supplementary Materials, Section 3: Robustness Checks on Mantel Test}\label{sm_mantel}
To assess the robustness of our findings, we have conducted three complementary sensitivity analyses examining the stability of our results under different data perturbations and alternative measurement approaches.

\begin{figure*}[!ht]
    \centering\includegraphics[width=\textwidth]{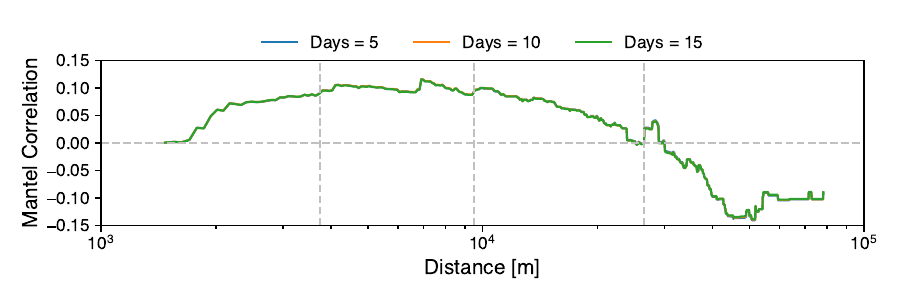}
    \caption{\textbf{Robustness analysis of Mantel correlogram under incomplete data scenarios.}
    Mantel correlogram based on incomplete datasets (with 5, 10, and 15 randomly removed days). The standard deviation is not shown, as it is negligible ($<10^{-4}$).
    }
    \label{fig_sm_mantel_random}
\end{figure*}

First, we have evaluated the resilience of our findings to incomplete data scenarios by simulating random data gaps. We have systematically removed N days from the original dataset ($N = 5$, $10$, $15$) and repeated this process $100$ times for each removal level. 
For each perturbed dataset, we have performed the complete analysis pipeline described in the main manuscript and compared the resulting Bray-Curtis dendrograms with spatial clustering patterns using the Mantel test. 
The Mantel correlograms displayed in Figure \ref{fig_sm_mantel_random} demonstrate remarkable consistency across all perturbation scenarios, with no visually distinguishable differences between the various removal levels or when compared to the main results. 

\begin{figure*}[!ht]
    \centering\includegraphics[width=\textwidth]{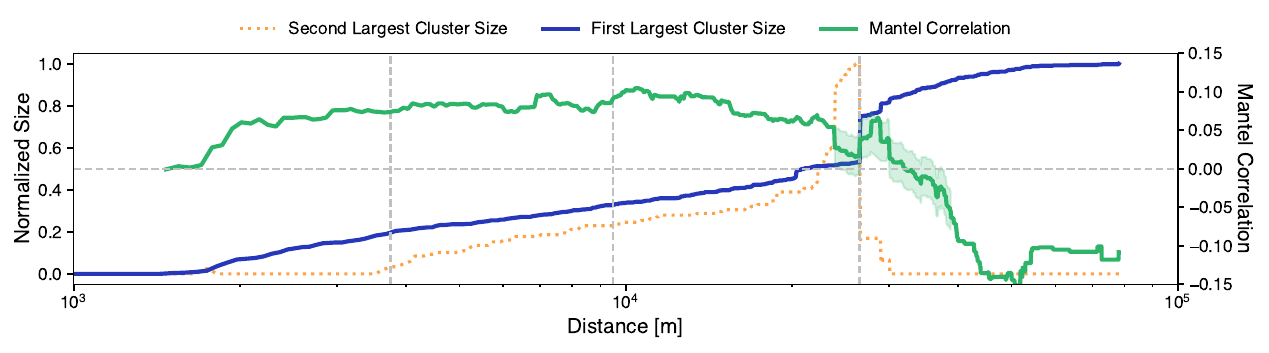}
    \caption{\textbf{Robustness analysis of Mantel correlogram using cumulative offered volume normalized by population.}
    Mantel correlation analysis (green) examines the relationship between spatial proximity and product composition similarity across multiple distance scales. The shaded area indicates $p$-values greater than $0.05$. For reference, the fraction of nodes in the largest spatial cluster, normalized by its maximum value (blue line), and the second-largest cluster (orange line) are shown. Vertical lines correspond to distances of $4000$ km, $9500$ km, and $26500$ km.
    }
    \label{fig_sm_mantel_count}
\end{figure*}

Second, we have examined whether our findings depend on the specific metric used to quantify product abundance. 
Instead of measuring abundance as the total number of days a product has been put on sale, we have computed the cumulative sum of individual volumes during the observation period. 
The results, presented in Figure \ref{fig_sm_mantel_count}, reveal the same pattern observed in the main analysis. 
However, this alternative metric produces a sharper transition from positive to negative correlations across distance classes, along with a narrower range of statistically irrelevant values.

\begin{figure*}[!ht]
    \centering\includegraphics[width=\textwidth]{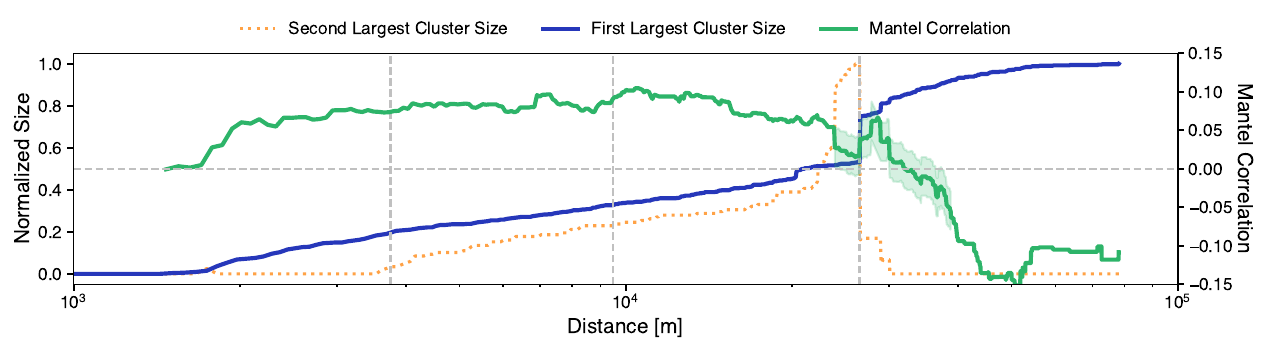}
    \caption{\textbf{Robustness analysis of Mantel correlogram using the cumulative sum of volume.}
    Mantel correlation analysis (green) examines the relationship between spatial proximity and product composition similarity across multiple distance scales. The shaded area indicates $p$-values greater than $0.05$. For reference, the fraction of nodes in the largest spatial cluster, normalized by its maximum value (blue line), and the second-largest cluster (orange line) are shown. Vertical lines correspond to distances of $4000$ km, $9500$ km, and $26500$ km.
    }
    \label{fig_sm_mantel_density}
\end{figure*}

Third, we have normalized the product abundance, described in previous paragraph, by dividing the total volume by the corresponding population of each store's municipality to account for potential demographic effects on product distribution patterns. 
As shown in Figure \ref{fig_sm_mantel_density}, this population-adjusted analysis preserves the core spatial clustering structure identified in previous results, confirming that our findings are not confounded by variations in local population density.
In both cases, the PERMANOVA results ($F = 5.1$, $p = 0.001$) demonstrate that spatial clusters correspond to distinct product composition strategies.
\pagebreak[2]
\clearpage
\section*{Supplementary Materials, Section 4: Robustness Checks on Bipartite Network}\label{sm_bipartite}

\begin{figure*}[!ht]
    \centering\includegraphics[width=\textwidth]{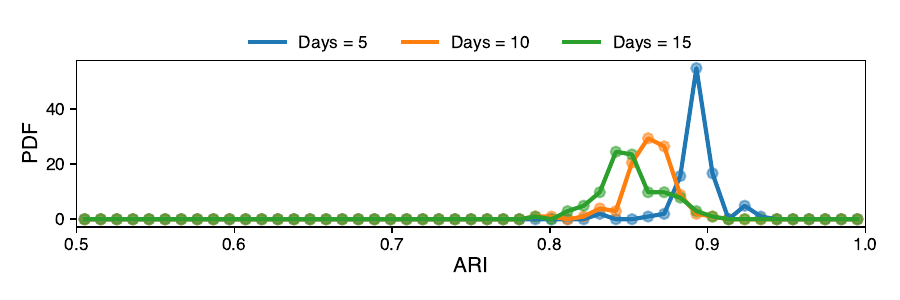}
    \caption{\textbf{Robustness analysis of network filtering under incomplete data scenarios.}
    The probability density functions show the distribution of Adjusted Rand Index scores comparing community structures derived from incomplete datasets (with 5, 10, and 15 randomly removed days) against the complete dataset.
    }
    \label{fig_sm_bipartite}
\end{figure*}

To assess the robustness of our bipartite approach, we have conducted a sensitivity analysis by simulating incomplete data scenarios. 
We have randomly removed $N$ days from the original dataset (repeated 100 times) and constructed the weighted bipartite network using the reduced data, applying the same RCA and BiCM filtering procedures described in the main manuscript. 
The robustness has been evaluated by comparing the community structure of the store projection network obtained from the incomplete data with that derived from the complete dataset. 
We have quantified this comparison using the Adjusted Rand Index (ARI) score between the detected communities, with results presented in Figure \ref{fig_sm_bipartite}. 
The analysis shows that incomplete data scenarios produce comparable results, consistently identifying three main communities across all tested conditions, with an average ARI score of $0.85$ between the 15-day removal scenario and the complete dataset, demonstrating the robustness of our methodology.
\begin{figure*}[!hb]
    \centering\includegraphics[width=\textwidth]{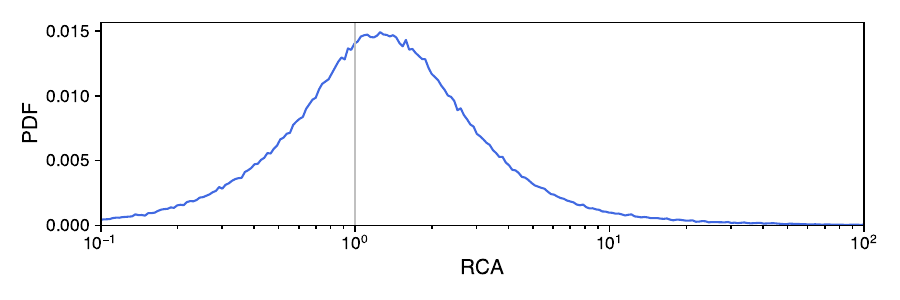}
    \caption{\textbf{Distribution of Revealed Comparative Advantage.}
    Distribution of the Revealed Comparative Advantage within the bipartite network presented in the manuscript, with the vertical grey line indicating the applied filter at $\text{RCA} = 1$.  
    }
    \label{fig_sm_rca}
\end{figure*}
\end{document}